# Solid-state dewetting instability in thermally-stable nanocrystalline binary alloys

Jennifer D. Schuler [a, b], Guild Copeland [b], Khalid Hattar [b], Timothy J. Rupert [a], Samuel A. Briggs [b, c, *]

[a] Department of Materials Science and Engineering, University of California, Irvine, CA 92697, USA
[b] Sandia National Laboratories, PO Box 5800, Albuquerque, NM 87185, USA
[c] School of Nuclear Science and Engineering, Oregon State University, Corvallis, OR 97331, USA

*E-mail: Samuel.Briggs@oregonstate.edu

**Abstract:**

Practical applications of nanocrystalline metallic thin films are often limited by instabilities. In addition to grain growth, the thin film itself can become unstable and collapse into islands through solid-state dewetting. Selective alloying can improve nanocrystalline stability, but the impact of this approach on dewetting is not clear. In this study, two alloys that exhibit nanocrystalline thermal stability as ball milled powders are evaluated as thin films. While both alloys demonstrated dewetting behavior following annealing, the severity decreased in more dilute compositions. Ultimately, a balance may be struck between nanocrystalline stability and thin film structural stability by tuning dopant concentration.

**Graphical Abstract**

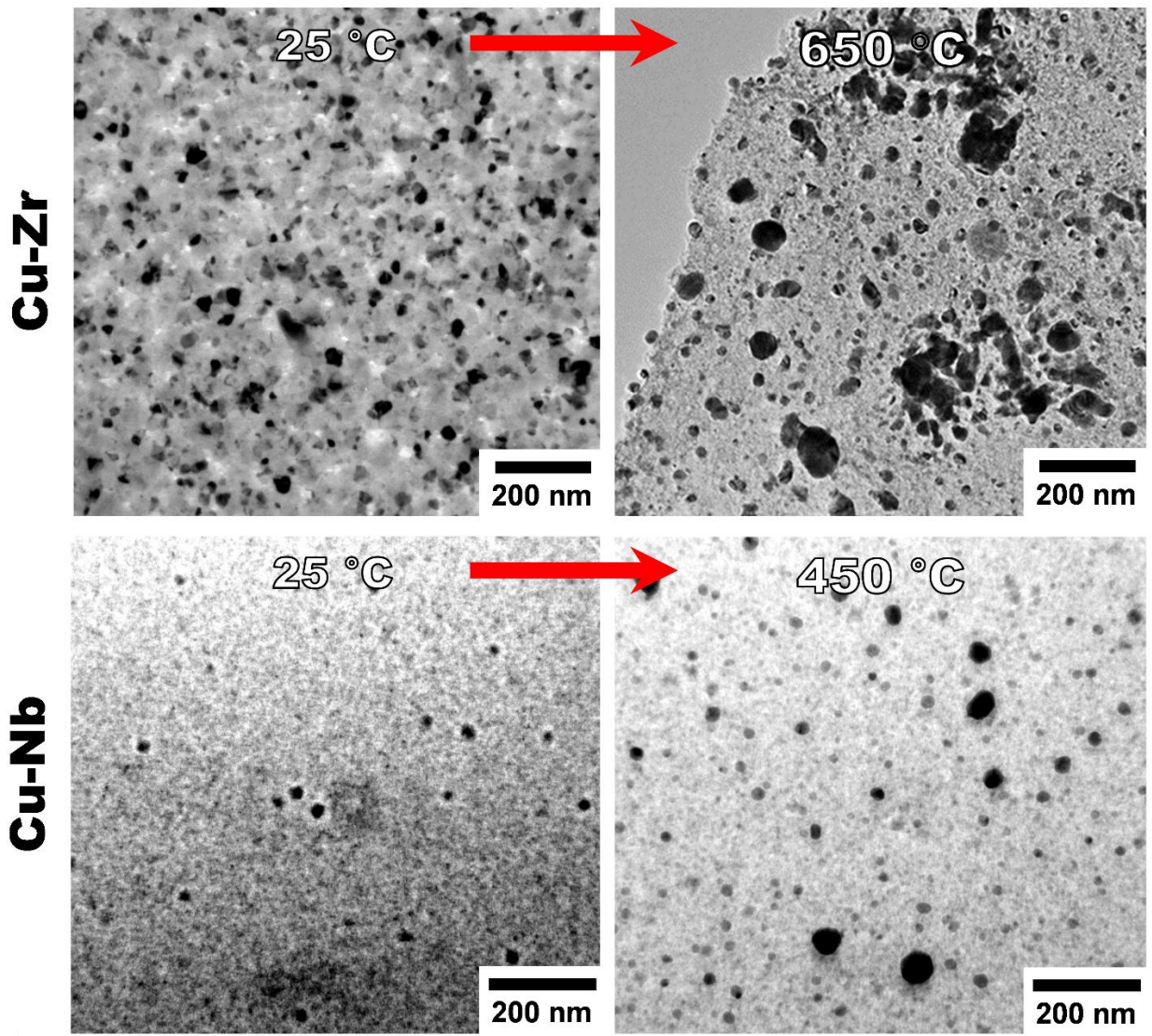

Nanocrystalline metals, defined as having an average grain size less than 100 nm, are desirable in thin film applications such as flexible electronics [1], microelectromechanical systems [2], and coatings [3], but are subject to high stresses that can lead to micro-cracking and delamination [4]. Additionally, thin films can experience solid-state dewetting [5, 6]. In order to minimize surface and interfacial energies, metastable thin films can rupture and collapse into particles with a shape dependent on the particle-substrate contact angle and Wulff construction at temperatures significantly below the melting temperature of the bulk phase [7]. Solid-state dewetting is attributed to enhanced diffusion and grain boundary grooving in polycrystalline films in which the groove grows until reaching the substrate, forming a hole which allows for grain breakaway and eventual coalescence with other dewetted regions [6, 8, 9]. Increasing the film thickness [10], decreasing surface roughness [11], tuning deposition conditions [12, 13], and substrate type selection [13-15] are all traditionally used to suppress solid-state dewetting.

In addition to solid-state dewetting, nanocrystalline metal thin films are also vulnerable to grain growth and thus a loss of the intended desirable properties generated by the fine grain structure, such as increased strength [16, 17], wear resistance [18-20], and corrosion resistance [21]. Methods to create thermally-stable nanocrystalline alloys involve selective alloying to access kinetic and thermodynamic grain growth limiting mechanisms [22-31]. Kinetic stabilization generally involves alteration of the grain boundary mobility through solute drag and Zener pinning caused by particle dispersions [32-38], which has been observed for Cu-Ta [38, 39] and Cu-Nb [40]. Thermodynamic stabilization methods seek to reduce the root driving force for grain growth by reducing the grain boundary energy through dopant segregation to the grain boundary region [23, 24, 26, 27, 41-44], which has been observed for a variety of systems including Ni-W [45], W-Ti [46], Fe-Zr [47], and Cu-Zr [48].

Kapoor and Thompson [49] recently reviewed how dopant segregation can be used to tailor thin film microstructures for a multitude of alloys, including Fe-Pt [50], Cu-Ni [51], Fe-Cr [52], and W-Ti [53]. Features such as the film stress, grain boundary distribution, texture, and grain size of the alloy films were all affected by dopant segregation. Sharma et al. [54] also observed that anisotropic metastable phases in thin films can drastically impact dewetting kinetics through suppression of grain boundary grooving. Selective alloying that alters grain boundary energy and mobility as well as thin film microstructural characteristics may also impact grain boundary grooving and solid-state dewetting. In this study, two binary alloys that have demonstrated nanocrystalline grain size thermal stability in ball milled powder forms are evaluated for their thermal stability as thin films, where the stability of Cu-Nb is primarily kinetic and the stability of Cu-Zr has a substantial thermodynamic component. We find solid-state dewetting can overpower both kinetic and thermodynamic stabilization from dopants, implying that additional considerations may be necessary to balance the desired nanocrystalline grain size and thin film structural stability.

Nanocrystalline Cu-Nb films with compositions of Cu-4 at.% Nb, Cu-8 at.% Nb, Cu-21 at.% Nb, and Cu-46 at.% Nb (hereafter referred to as Cu-4Nb, Cu-8Nb, Cu-21Nb, and Cu-46Nb, respectively) were produced through magnetron co-sputtering using an Ar plasma with a Kurt J. Lesker Lab 18 modular thin film deposition system operated at room temperature using a pressure of 12 mTorr, power of 400 W for Cu, and variable powers for Nb to achieve the desired film composition. Nanocrystalline Cu-5 at.% Zr (hereafter referred to as Cu5Zr) thin films were produced using an Ulvac JSP 8000 metal deposition sputter tool, with deposition performed at room temperature using an Ar pressure of 1.5 mTorr, Cu power of 150 W, and a Zr power of 75 W. The films were deposited to a thickness of approximately 50 nm to maintain electron transparency onto either polished, single-crystal (100) orientation NaCl substrates or $Si_3N_4$ window transmission electron microscopy (TEM) grids. Specimens deposited on NaCl substrates were subsequently placed as freestanding thin films onto Mo mesh grids through dissolution of the substrate. Bright-field TEM and *in situ* TEM annealing experiments were performed either on a JEOL Grand ARM300CF TEM/STEM operating at 300 kV or a JEOL 2100(HT) at 200 kV using a Gatan 652 double-tilt heating holder with a ramp rate of 20 °C/min to a maximum temperature of 450 °C for Cu-Nb and 50 °C/min to 650 °C for Cu-Zr. Energy-filtered TEM (EFTEM) was performed on the JEOL Grand ARM300CF for Cu-Zr samples using the jump ratio method with the L-edge threshold energies for Cu and Zr. Automated Crystal Orientation Mapping (ACOM) with precession electron diffraction (PED) was performed on the JEOL 2100(HT) using a 5 nm step size, precession angle of 0.375°, and 10 precessions per frame. Focused ion beam (FIB) channeling contrast imaging and energy-dispersive X-ray spectroscopy (EDS) were performed using an FEI Quanta 3D FEG Dual-Beam scanning electron microscope (SEM) operating at 30 kV. EDS data was obtained at 30 kV, and Cu K$\alpha$, Nb L$\alpha$, and Zr L$\alpha$ peaks were used for compositional analysis using the Inca software suite from Oxford Instruments.

The bright-field TEM images in Figs. 1(a)-(c) show the progression of solid-state dewetting in Cu-46Nb deposited onto $Si_3N_4$ window TEM grids during *in situ* TEM annealing at 25 °C, 215 °C, and 450 °C, respectively. The arrows in Figs. 1(b) and (c) track selected particles that grow as temperature is increased. ACOM phase analysis of dewetted particles is presented in Fig. 1(d) for Cu-4Nb, which also exhibited solid-state dewetting when annealed at 500 °C for 1 h, where the face centered cubic (fcc) Cu phase is shown in red. The inset in Fig. 1(d) shows the associated bright-field TEM image. The ACOM map shows that the large dewetted particles are a Cu-rich fcc phase. Any Nb precipitates, if present, were too small to be visible in the ACOM map. Cu-Nb freestanding thin films on Mo mesh grids subjected to heating (not shown here) also experienced similar solid-state dewetting. Cu and Nb are mostly immiscible [55], meaning that an alloy containing these constituents will decompose into each element in its pure form. Ball milled bulk Cu-Nb has demonstrated nanocrystalline stability up to 1000 °C in previous studies due to kinetic stabilization caused by Nb precipitation and Zener pinning [40, 55, 56]. Cu-Nb as a thin film was instead structurally altered by the formation and growth of dewetted Cu particles, where the dewetted particles increased in total area with increasing temperature.

Bright-field TEM images in Figs. 2(a) and (b) at 25 °C and 650 °C, respectively, show similar solid-state dewetting in Cu-5Zr freestanding thin films on Mo mesh grids, where large dewetted particles are observed. The insets show the selected area electron diffraction patterns for each condition where fcc Cu rings (solid red lines) were present at 25 °C in addition to $Cu_2O$ (dashed green lines), which may be due to oxidation from trace O during vacuum annealing or from the deposition and film preparation processes. Upon heating, only the fcc Cu rings are observed. It is possible that Cu oxides in the film can reduce to Cu during vacuum annealing and then serve as initiation points for dewetting [57]. Future work is needed in order to correlate sites where dewetting initiates and the local film condition. Fig. 2(c) shows the post-anneal EFTEM composition map collected from another region of the annealed film in Fig. 2(b), where large dewetted Cu particles are present. Ball milled Cu-Zr has been found to remain nanostructured even after a week at 98% of its solidus temperature [48], which was primarily attributed to thermodynamic stabilization caused by dopant segregation and grain boundary complexion transformations. Instead of a thermodynamically stabilized material, the thin Cu-5Zr film collapsed into large dewetted Cu particles.

In order to better understand the impact of dopant concentration on the film grain size stabilization and dewetting characteristics, a variety of Nb concentrations were evaluated. Figs. 3(a)-(h) show the microstructural evolution of the Cu-4Nb, Cu-8Nb, Cu-21Nb, and Cu-46Nb freestanding thin films on Mo mesh grids after annealing under vacuum at 500 °C for 1 h. Figs. 3(a)-(d) show the films in the as-deposited state, while Figs. 3(e)-(h) show the films after being annealed. The insets show the selected area electron diffraction patterns for each condition. The fcc Cu rings are present in all of the thin films (solid red lines). Some $Cu_2O$ phase was also again detected (dashed green lines). The body centered cubic (bcc) Nb phase (dashed blue lines) was only detected in the Cu-21Nb and Cu-46Nb films, which may be due to dewetting shifting local dopant concentrations or the high global Nb content in the as-deposited state coupled with local film inhomogeneities. Fig. 3(i) shows the percentage of area covered by dewetted particles after annealing. The initial percentage of area covered by dewetted particles before annealing (the as-deposited state) was approximately zero for all Cu-Nb alloys as no particles were observed. The Cu-4Nb (Figs. 3(a) and (e)) and Cu-8Nb (Figs. 3(b) and (f)) films remained nanocrystalline with limited dewetting, where only 3% and 7%, respectively, of each inspected area was comprised of dewetted particles after annealing. The Cu-21Nb (Figs. 3(c) and (g)) and Cu-46Nb (Figs. 3(d) and (h)) films exhibited increased heterogeneity in the as-deposited state, and formed large dewetted Cu particles following annealing, where 12% and 38%, respectively, of each inspected area was consumed by dewetted particles. While increasing Nb concentrations correlate to enhanced kinetic stabilization against grain growth due to Nb precipitation [55], solid-state Cu dewetting still occurs and actually becomes more intense with increasing Nb concentration. All Cu-Nb and Cu-Zr thin films on either $Si_3N_4$ window or Mo mesh TEM grids formed extensive holes that grew as dewetting progressed.

The impact of film thickness on solid-state dewetting was also investigated by increasing the thickness of the Cu-Zr film. Fig. 4 shows FIB channeling contrast images of a 2 µm-thick version of the Cu-5Zr thin film that was presented in Fig. 2, with this thicker film deposited onto a Cu substrate polished to a mirror finish. The thicker film was then annealed at 900 °C to induce a complexion transformation at some boundaries to an amorphous intergranular film that contributes to enhanced nanocrystalline grain size stability, as shown in our prior work in Refs. [48, 58, 59]. In FIB channeling contrast images, differences in contrast refer to different crystalline orientations. In Fig. 4(a), large dewetted particles are shown containing micrometer-scale grains orders of magnitude larger than the grains of the nanocrystalline Cu-Zr film beneath the dewetted particles. Fig. 4(b) shows a FIB channeling contrast image of a cross-section from a separate trench. Consumption of Cu by the dewetted particles makes the thin film porous with a higher Zr concentration than the as-deposited state. The percentage of the cross-sectional area replaced by pores was measured from the thin film region underneath the large dewetted particle in Figure 4(b) to be ~15%. The inset in Fig. 4(b) shows a magnified image of the porous film, where a single pore is indicated by a dashed yellow circle. SEM-EDS elemental compositions for each region are also presented in Fig. 4(b), where the dewetted particle has <1 at.% Zr and the porous enriched film has 14 at.% Zr, despite the global as-deposited film composition of 5 at.% Zr [60]. Trace amounts of Zr detected in the dewetted particle may be attributed to the adjacent sputtered film being captured by the electron beam interaction volume or the small amount of Zr solid solubility in Cu. Even though larger film thicknesses correlate to suppressed solid-state dewetting [10], many dewetted Cu particles were still observed.

Solid-state dewetting theory has been well established within the context of single element systems, but binary alloys pose a more complex structural and chemical environment that adds new dimensions to the evolution of dewetted particles [61-64]. Grain boundary grooving and subsequent solid-state dewetting has been attributed to atomic diffusion along grain boundaries and interfaces that transports mass away from holes towards hillocks that protrude from the film surface [6]. The solid-state dewetting observed in this study may also be due to Cu self-diffusion along grain boundaries. Despite variations in film thickness, substrate, annealing temperature, and dopant interactions, the degree of dewetting was observed to be most impacted by the dopant concentration. While the Cu-5Zr films studied here demonstrated extensive solid-state dewetting, Cu-Zr with compositions ranging from 0.3 to 1.2 at.% Zr have been previously shown to not exhibit solid-state dewetting [65]. Lower dopant concentrations correlate to decreased surface roughness and lattice strain which can suppress dewetting, as has been shown for Cu-Ag [11] and Cu-Mo [66]. The enhanced dewetting observed in the Cu-Nb films with higher Nb concentrations may be due to increased surface roughness and lattice strains as Nb is added. In addition, LaBarbera [12] reasoned that grain boundary grooving can play a more dominant role in advancing solid-state dewetting in nanocrystalline alloys due to their increased grain boundary volume fraction [67]. Future work probing compositional variations and atomic diffusion at grain boundaries and dewetted regions is needed in order to fully understand the solid-state dewetting mechanisms of these thin films.

Dilute dopant concentrations that suppress solid-state dewetting also influence the stabilization of the grain boundaries. A grain boundary groove in a polycrystalline film can be described by the surface dihedral angle of the groove, $\psi_s$ [7, 63, 68]:

$$\gamma_{gb} = 2\gamma_{surf} \cos\left(\frac{\psi_s}{2}\right) \tag{1}$$

where $\gamma_{gb}$ is the grain boundary energy, and $\gamma_{surf}$ is the isotropic surface energy. Additionally, groove geometry changes as a function of grain boundary energy and character, where higher energy grain boundaries form deeper grooves than lower energy grain boundaries [68-72]. As a result, solid state dewetting is more likely to occur at high energy grain boundaries. The thermodynamic grain size stabilization pathway utilizes grain boundary dopant segregation to lower the grain boundary energy [22, 28], which may also contribute to forming shallower grooves and suppressing solid-state dewetting. Future work to measure and compare the dihedral angles of grain boundary grooves in thermodynamically versus kinetically stabilized films would improve the field's understanding of how dewetting mechanisms are impacted by different dopants.

Kinetic stabilization of the grain size through grain boundary pinning may also suppress grain boundary grooving by enacting a pinning pressure to counteract groove advancement. The Zener pinning pressure $P_z$ can be described by [36]:

$$P_z = \frac{3 \cdot F_v \cdot \gamma_{gb}}{2r} \qquad (2)$$

where $F_v$ is the volume fraction of randomly distributed spherical particles, $r$ is the particle radius, and $\gamma_{gb}$ is the grain boundary energy. Since the pinning pressure is a function of the grain boundary energy, it may be possible to apply principles from Eqn. 1 and find the pinning pressure dependence on groove geometry, where a critical groove dihedral angle must first be achieved to overcome the Zener pinning pressure and allow for groove advancement. Future work to calculate the Zener pinning pressure of the Cu-Nb films presented in this study would provide critical insight in understanding its connection to grain growth and solid-state dewetting. Ultimately, selective alloying for thermodynamic or kinetic grain size stabilization may limit grain boundary grooving and suppress the initiation of solid-state dewetting, enhancing both the grain size stability and thin film structural stability.

While low dopant concentration suppresses solid-state dewetting by decreasing surface roughness and lattice strains, grain size thermal stability through both thermodynamic and kinetic mechanisms are dependent on dopant concentration, which has been observed for Fe-Zr [47], Ni-W [45], and Cu-Nb [40, 55]. More severe solid-state dewetting has been observed in pure Cu [10, 11, 13, 15, 65] and high dopant concentration Cu alloy thin films compared to Cu alloys with more dilute concentrations. This suggests that a balance must be struck in order to avoid both grain growth and solid-state dewetting, where sufficient dopant is present to retain nanocrystallinity while minimizing surface roughness and lattice strain to inhibit solid-state dewetting, creating a film with optimal grain size and structural stability. Too little dopant can make an alloy unstable against grain growth, while too much can make it unstable against dewetting.

In summary, two alloy systems that have exhibited nanocrystalline thermal stability through kinetic (Cu-Nb) and thermodynamic (Cu-Zr) stabilization as ball milled powders have been evaluated for their thermal stability as sputter-deposited thin films. In both cases, microstructural evolution was dominated by the formation of Cu particles through a solid-state dewetting mechanism, damaging the film structural stability. Dopant concentration was observed to have the greatest impact on solid-state dewetting, where more dilute dopant concentrations limit the dewetting behavior. The same thermodynamic and kinetic mechanisms leveraged to stabilize nanocrystalline alloys may also contribute to enhanced thin film structural stability by limiting grain boundary grooving, a primary mechanism behind solid-state dewetting. While more dilute alloys limit dewetting, low dopant concentrations can also result in decreased nanocrystalline stability, implying that ideal metallic thin film compositions must balance nanocrystalline and thin film structural stabilities.

**Acknowledgements:**

JDS and TJR acknowledge the U.S. Department of Energy, Office of Basic Energy Sciences, Materials Science and Engineering Division under Award No. DE-SC0014232, and the U.S. Department of Energy, Office of Science, Office of Workforce Development for Teachers and Scientists, Office of Science Graduate Student Research (SCGSR) program. The SCGSR program is administered by the Oak Ridge Institute for Science and Education for the DOE under contract number DE-SC0014664. SAB and KH were supported by the U.S. Department of Energy, Office of Basic Energy Sciences, Materials Science and Engineering Division under FWP 18-013170. This work was performed, in part, at the Center for Integrated Nanotechnologies, an Office of Science User Facility operated for the U.S. Department of Energy (DOE) Office of Science. Sandia National Laboratories is a multimission laboratory managed and operated by National Technology & Engineering Solutions of Sandia, LLC, a wholly owned subsidiary of Honeywell International, Inc., for the U.S. DOE's National Nuclear Security Administration under contract DE-NA-0003525. The views expressed in the article do not necessarily represent the views of the U.S. DOE or the United States Government. Analytical TEM work was performed at the UC Irvine Materials Research Institute (IMRI). SEM and FIB work was performed at IMRI using instrumentation funded in part by the National Science Foundation Center for Chemistry at the Space-Time Limit (CHE-0802913).

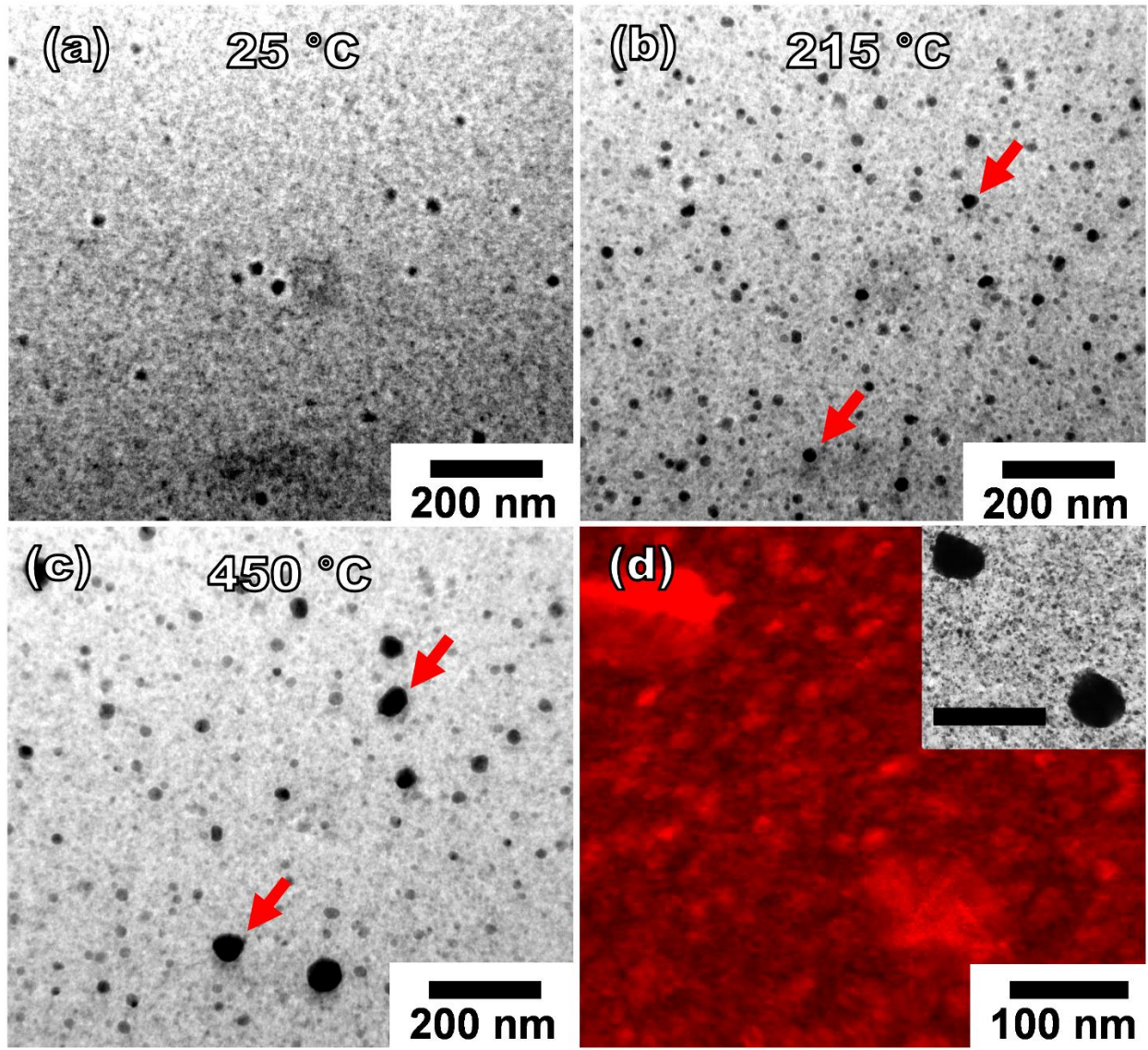

**Fig. 1**. (a)-(c) Evolution of solid-state dewetting in Cu-46 at.% Nb with increasing temperature during *in situ* heating experiments. The arrows in (b) and (c) track the growth of selected dewetted particles. (d) Phase analysis of Cu-4 at.% Nb after being annealed at 500 °C for 1 h using precession electron diffraction, where fcc Cu is red. The inset shows the associated bright-field TEM image, where the scale bar is 250 nm.

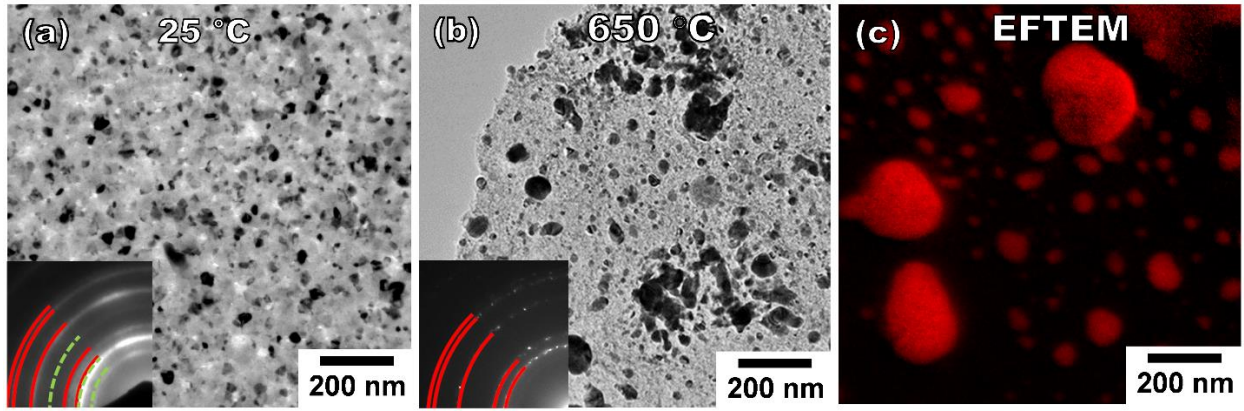

**Fig. 2**. Evolution of solid state dewetting in Cu-Zr with increasing temperature during *in situ* heating experiments. The bright-field TEM image in (a) shows the film in the as-deposited state at 25 °C, while (b) shows the film after annealing at 650 °C. The insets show the associated selected area electron diffraction patterns for each condition, where fcc Cu is indicated by solid red lines and the $Cu_2O$ phase appears as dashed green lines. (c) Elemental mapping using EFTEM on a separate region of the annealed film, where Cu appears red.

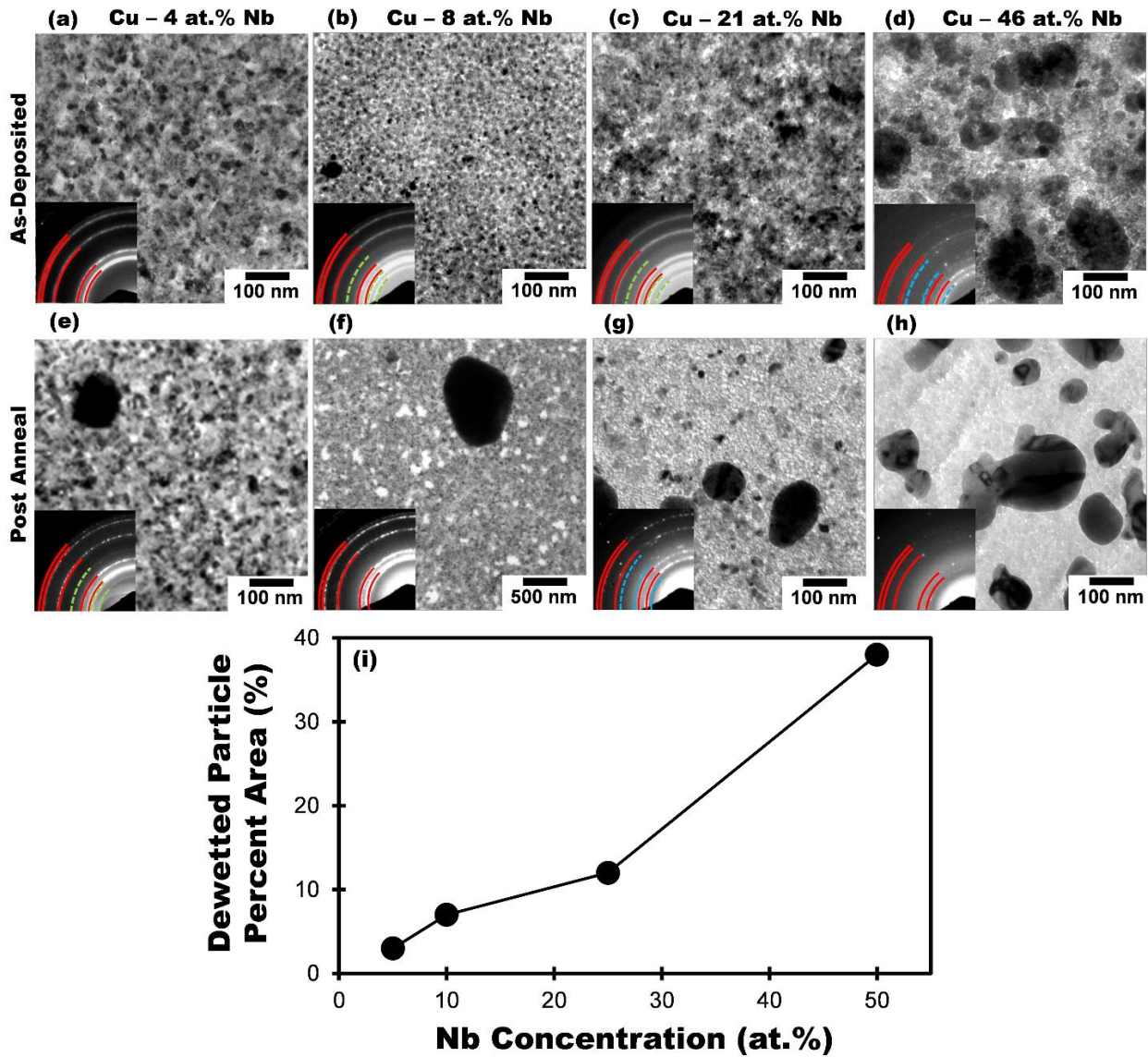

**Fig. 3**. Bright-field TEM images showing the microstructural evolution of Cu-Nb alloys as a function of Nb concentration. The microstructure of each film in the as-deposited state is shown for (a) Cu-4 at.% Nb, (b) Cu-8 at.% Nb, (c) Cu-21 at.% Nb, and (d) Cu-46 at.% Nb. The microstructure of each film after annealing at 500 °C for 1 hour is presented in (e)-(h). The insets show the associated selected area electron diffraction patterns for each condition, where fcc Cu phase is indicated by solid red lines, the $Cu_2O$ phase by dashed green lines, and bcc Nb by dashed blue lines. (i) the percentage of area covered by dewetted particles after annealing.

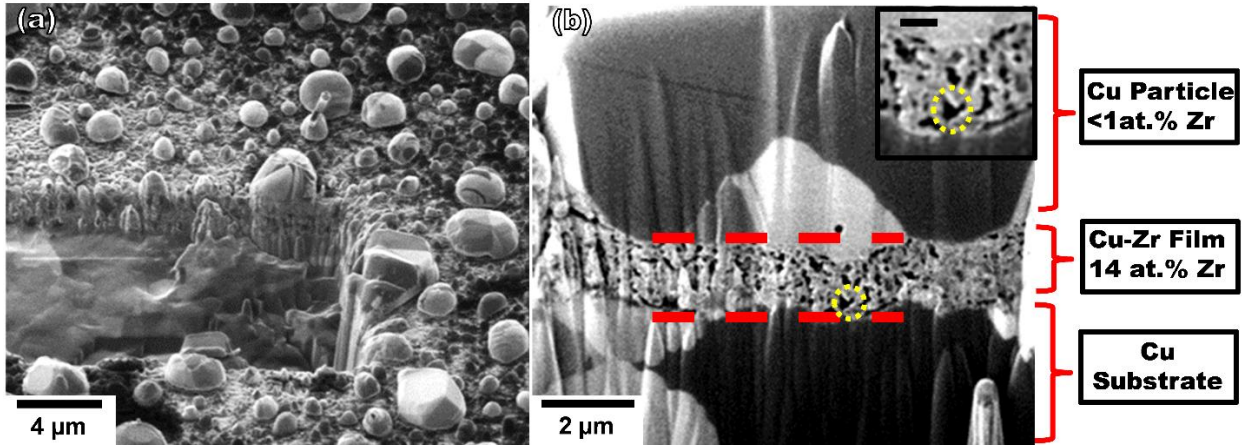

**Fig. 4**. FIB channeling contrast imaging of the Cu-5 at.% Zr thin film deposited on a Cu substrate after being annealed at 900 °C. (a) Dewetted Cu particles containing large grains are dispersed across the film surface. (b) A separate trench where the porous remains of the sputtered film can be seen beneath a dewetted particle. The labels in (b) show the associated compositional data for the film and particle. The inset in (b) shows a magnified image of the porous film, where a single pore is indicated by a dashed yellow circle. The scale bar in the inset is 600 nm.